\documentclass[%
 reprint,
 amsmath,amssymb,
 aps,
]{revtex4-2}

\usepackage{graphicx}
\usepackage{dcolumn}
\usepackage{bm}

\begin{document}

\preprint{APS/123-QED}

\title{Machine learning models for atom-diatom reactions across isotopologues}

\author{Daniel Julian}
\author{Rian Koots}
\author{Jes\'us P\'erez-R\'ios}

\affiliation{Department of Physics and Astronomy, Stony Brook University, Stony Brook, New York, 11794-1901, United States
}%

\date{\today}

\begin{abstract}
 This work shows that feed-forward neural networks can predict the final ro-vibrational state distributions of inelastic and reactive processes of the reaction of Ca + H$_2$$\rightarrow$ CaH + H in the hyper-thermal regime, relevant for buffer gas chemistry. Furthermore, these models can be extended to the isotopologues of the reaction involving deuterium and tritium. In addition, we develop a neural network model that can learn across the chemical space based on the isotopologues of hydrogen. The model can predict the outcome of a reaction whose reactants have never been seen. This is done by training on the Ca + H$_2$ and Ca + T$_2$ reactions and subsequently predicting the Ca + D$_2$ reaction. 
\end{abstract}

\maketitle


\section{Introduction }

Artificial intelligence and machine learning are becoming ever more prominent in the big-data era, and this is certainly true in the physical and chemical sciences. In the study of physical chemistry, powerful data-driven tools can be used to make predictions about the outcomes of chemical reactions~\cite{Meuwly2021,chemistry,chemistry2,chemistry3,Andrade2023}, the spectroscopic properties of molecules~\cite{spectroscopy1,spectroscopy2,spectroscopy3,Ibrahim2024,Liu2021,bindingPs}, the discovery of new materials~\cite{materials,materials2,materials3}, development of quantum technologies~\cite{QuantumTechnolgoies} and the design of new and efficient experimental protocols~\cite{ImageML,LightML,BECML}. In each scenario, the driving force is the same: finding universal models to ensure reliable predictions without performing costly computations.

Universal machine learned models capable of making predictions for many chemical reactions would revolutionize chemical engineering and manufacturing by unlocking the optimal conditions under which reactions can be produced. Motivated by this possibility, the physical chemistry community is working on developing universal data-driven chemical reaction predictors. However, the task is heroic due to the vastness of the chemical space. On the other hand, it is possible to focus on small, fully controllable systems to unleash the predictive power of machine learning: atom-diatom collisions, which, at low temperatures, find applications in atomic and molecular physics~\cite{Timur,Morita}. In that endeavor, there have been several efforts toward predicting atom-diatom collisions in the hyperthermal regime~\cite{Arnold2020,Arnold2022,COO,Dissociation,Koner2019,Panesi2023,Veliz2022,Gu2023,wang2024highenergy}. Specifically, it has been shown that neural networks (NN) can efficiently predict the state-to-state cross sections and the product state distributions. However, most of these studies focused on the reactive pathway, and only recently, a combined study was performed: inelastic plus reactive processes for O + CO reactions have been reported~\cite{COO}, albeit in this reaction, the reactive process only accounts for atom exchange. In addition, and more importantly, the machine learning models developed can only make predictions on previously seen reactions; i.e., the machine learning models are exposed to data for the same reaction to be predicted, which constrains the universality of the models.




In this work, we show that it is possible to predict the product state distributions of an atom-diatom reaction across the isotopologue space via deep learning. Precisely, we predict the final ro-vibrational state distributions of the reaction Ca + H$_2$ $\rightarrow$ CaH + H in the hyperthermal regime, and those related isotopologues involving the hydrogen isotopes. We show that a reduced featurization is suitable for training independent models capable of individually predicting each isotopologue of the reaction. Furthermore, this featurization can be used for models predicting the final ro-vibrational state distributions of inelastic processes. To move towards a more universal atom-diatom reaction predictor, we develop a model capable of generalizing between the isotopologues of the reaction by taking into account the isotopic effects of the reaction using an extended featurization. The paper is organized as follows: Section~\ref{method} presents the fundamentals of quasi-classical trajectory (QCT) calculations and the main characteristics of the NN models; Section~\ref{results} presents the main results based on our NN models and Section~\ref{conclusions} contains a summary and main conclusions derived from this work. 

\section{Methodology}\label{method}
We focus on Ca+H$_2(v_i,j_i)$ collisions in the hyper-thermal regime (collision energies $\gtrsim 1000$~K) where $(v_i,j_i)$ denotes the initial ro-vibrational sate of the hydrogen molecule. In this regime, the collision shows three possible reaction products, characterized as:

\begin{itemize}

\item Reaction: Ca+H$_2(v_i,j_i)$ $\rightarrow$ CaH$(v_f,j_f)$ + H. The CaH diatom is formed, where $(v_f,j_f)$ stands for the ro-vibrational state of the product molecule. This reaction could be used to efficiently produce CaH, one of the molecules of interest in cold and ultracold chemistry~\cite{CaH,CaH2}, in a controllable manner, such as previous work showed~\cite{BufferGasChemistry}. 

\item Inelastic collision:  Ca+H$_2(v_i,j_i)$ $\rightarrow$  Ca+H$_2(v_f,j_f)$. The reactants and products are the same, but they appear in different internal states. 

\item Dissociation: Ca+H$_2(v_i,j_i)$ $\rightarrow$ Ca + H + H. Here, all three atoms become unbounded and no product diatom is formed. 
\end{itemize}

\noindent
Moreover, we consider the isotopologues of the scattering process Ca+H$_2(v_i,j_i)$ with the deuterium and tritium molecules replacing hydrogen, as shown in Table~\ref{chem_eqs}.

\begin{table*}[t!]
    \centering
    \begin{tabular}{|c|c|c|}
    \hline
        Isotopologue & Reactive Process  & Inelastic Process  \\
         \hline
         Hydrogen & H$_2$($v_i$, $j_i$) + Ca $\rightarrow$ CaH($v_f$, $j_f$) + H & H$_2$($v_i$, $j_i$) + Ca $\rightarrow$ H$_2$($v_f$, $j_f$) + Ca   \\
         \hline
         Deuterium & D$_2$($v_i$, $j_i$) + Ca $\rightarrow$ CaD($v_f$, $j_f$) + D  & D$_2$($v_i$, $j_i$) + Ca $\rightarrow$ D$_2$($v_f$, $j_f$) + Ca  \\
         \hline
         Tritium & T$_2$($v_i$, $j_i$) + Ca $\rightarrow$ CaT($v_f$, $j_f$) + T  & T$_2$($v_i$, $j_i$) + Ca $\rightarrow$ T$_2$($v_f$, $j_f$) + Ca  \\
         \hline
    \end{tabular}
    \caption{The isotopologues of the reactive collision along with the processes that are investigated. It should be noted that no NN model was developed for the inelastic process of the tritium isotopologue.}
    \label{chem_eqs}
\end{table*}

The goal of our neural network (NN) models is to learn the final product state distributions of reactive and inelastic collisions, excluding dissociation since this process does not produce molecules. Therefore, we need data to train these models. In our case, training data is generated via QCT calculations, which yields reliable results in this energy regime~\cite{Arnold2020,COO,Panesi2023}.

\subsection{QCT calculations}

QCT calculations treat the nuclear dynamics classically in the potential energy described by the electrons of the system. In contrast, the initial conditions to evolve the classical equations of motion depend on the quantum state of the system via the Bohr-Sommerfeld quantization rule~\cite{truhlar1979}. Similarly, the outcome of the classical evolution can be expressed in a quantum manner using the same principle. QCT calculations are accurate to describe the dynamics of small systems at high collision energies, where many partial waves contribute to the scattering observables, and quantum effects are washed out~\cite{rios2020introduction}. 

In these simulations, the interaction potential between the atoms is specified, and many trajectories are run with randomized initial conditions (generated via the Monte Carlo method) over a range of impact parameters. The molecule is initialized in a ro-vibrational state corresponding to its energy spectrum $(v_i,j_i)$ and the atom is initialized beyond the interaction range of the molecule. The system is evolved classically using Hamilton's equations of motion, and the final ro-vibrational state is assigned semi-classically using the Bohr-Sommerfeld quantization rule. 
The final rotational number is given by~\cite{truhlar1979}
\begin{equation}
    j_f = -\frac{1}{2} + \frac{1}{2}\sqrt{1+4\frac{\vec{J_f}\cdot{\vec{J_f}}}{\hbar^2}},
\end{equation}
where $\vec{J_f}$ is the relative angular momentum of the molecule.
The final vibrational number for a molecule is given by~\cite{truhlar1979}
\begin{equation} \label{eq: vprime}
    v_f = -\frac{1}{2} + \frac{\sqrt{2\mu}}{\pi\hbar}\int_{r_-}^{r_+}\left[E'_{int} - V(r) - \frac{\hbar^2j_f(j_f+1)}{2\mu r^2}\right]^{\frac{1}{2}}dr,
\end{equation}
where $\mu$ is the reduced mass of the product molecule, $V(r)$ is the molecular potential energy as a function of $r$, the interparticle distance, and [$r_{-},r_{+}$] are the inner and outer turning points of the molecule, respectively. Both $v_f$ and $j_f$ are treated with the histogram binning method, where the above expressions are rounded to the nearest integer. 

In our calculations we assume a pair-wise interaction potential, in which each pair-term is described by a Morse potential: $V(r) = D_e\left(1-\exp^{(-\alpha(r-r_e))}\right)^2 - D_e$. The potential parameters for H$_2$ and CaH were obtained from~\cite{h2_par,h2_pec} and ~\cite{cah_pec,cah_par}, respectively. We have checked that adding a non-additive interaction does not change the shape of the product state distributions, and only changes them through an overall factor. For a given collision energy, we run 10$^4$ trajectories per impact parameter in the range of 0 - 3.75 a$_0$ sampled at equal intervals of 0.25 a$_0$; i.e., we run 160,000 trajectories per initial condition. The collision energies were in the range 5,000 K - 50,000 K. A limited number of initial rotational and vibrational quantum numbers were sampled given the tendency of the diatoms to dissociate under high ro-vibrational states. Vibrational quantum numbers were sampled in the range $v_i=$ [0, 9] and rotational numbers were sampled in $j_i=$ [0, 20]. A limited number of initial ro-vibrational states are explored because the lightest isotopologue of the reactant diatom, namely H$_2$, cannot support many ro-vibrational states before it dissociates.

\subsection{Targets}

The final product state distributions are discrete probability mass functions that give the distribution of the final rotational and vibrational states with the probability of either the reactive or inelastic process. The distributions are chosen in such a way that each of the rotational and vibrational distributions is marginalized over the other. Each initial state, characterized by the collision energy $E_c$, $v_i$, and $j_i$, produces 
a final rotational state distribution for $j_f$ and a final vibrational state distribution for $v_f$. Furthermore, for a given final rotational or vibrational distribution, the sum of probabilities over all quantum numbers gives the total probability of either the reactive or inelastic process. Thus, the NN models for the product state distributions aim to provide a prediction of the distribution of final states as well as the probability of the reactive or inelastic process for a given initial state.

\subsection{Featurization}
In this work we develop two main classes of models. The first class consists of models that make predictions for a particular isotopologue and process (reactive or inelastic) whose training data consists solely of that particular process. The second class consists of a model that was trained on two reactive processes, the hydrogen and tritium isotopologues of the reaction, that can make predictions across the isotopologues for the reactive process in Table \ref{chem_eqs}. In other words, the second model predicts a reaction that has never been exposed to the NN before, namely the deuterium isotopologue, thus exploring the chemical space. The first class does not require any information that distinguishes the hydrogen isotopes, whereas the second class has an expanded featurization that aptly differentiates the isotopologues.

 The featurization for the models includes information about the initial state and the reactant diatom. The model that makes predictions across isotopologues requires additional features, such as masses and select spectroscopic constants of the reactant diatom. For the first class of models, the featurization is as follows: the collision energy $E_c$ in K, the initial rotational quantum number $j_i$, the initial vibrational quantum number $v_i$, the internal energy of the diatom, the angular momentum of the diatom, the rotational energy of the diatom (with $v$ = 0), the vibrational energy of the diatom (with $j$ = 0), the vibrational time period $\tau$, the relative velocity, and the classical turning points of the diatom (these count as two separate features), following Ref.~\cite{Koner2019}. For the second class, which makes predictions across isotologues, the following additional features were included: the reduced mass of the reactants, the mass of the hydrogen isotope, the reduced mass of the product diatom, the harmonic frequency spectroscopic constant $\omega_e$ of the reactant diatom, the rotational spectroscopic constant $B_e$ of the reactant diatom, and the binding energy spectroscopic constant of the reactant diatom $D_0$. By extending the feature vector for the second class of model to inform about the isotopic effects of the collision, the model has become more general by learning multiple chemical reactions.

\subsection{Neural Networks}

 Neural networks (NN) are machine-learned models that fit a multitude of parameters to learn an underlying relationship presented in its training data. In this work, we use the simplest conception of a deep NN: a fully-connected feed-forward neural network with multiple hidden layers. Feed-forward NNs consist of layers, each having a certain number of neurons that act as perceptron units. These perceptron units perform a linear transformation on its input followed by a non-linear transformation, also known as the activation, to produce the neuron's output. First, an input vector consisting of the features of the model is fed to the first layer, and each neuron in the first layer produces an output. All of the outputs of the first layer's neurons are collected into another vector, which is fed into each neuron of the second layer. This process continues until the final layer is reached, and the vector of the outputs of the last layer are the predictions, often regarded as the target vector. The fact that each neuron's output is fed into each neuron of the following layer makes the model fully-connected. 
 
 In the learning process, an optimization algorithm, most of the time a variant of gradient descent, is used to fit the parameters of the model, which are the weights and biases in the linear transformations of the neurons. The learning process involves continually updating the model parameters to minimize a loss function, which measures the error in the model's predictions compared to what is actually present in the training data set. Though the parameters in the linear transformation the focus of the learning process, the non-linear activation functions are necessary for the model to learn complex relationships since they introduce the ability to learn non-linear relationships, which are common in virtually all applications. Aspects of the model, such as its architecture, learning rate, initialization scheme, optimization algorithm, loss function, and number of training iterations, are known as hyper-parameters. Hyper-parameters are not automatically fit and must be chosen in such a way, usually empirically, to best allow the model to learn from the training data without overfitting to it.

The training data for the NN models was generated via quasi-classical trajectory (QCT) simulations using the PyQCAMS software package \cite{PyQCAMS}. The results of the simulation for each initial condition were used to generate the final rotational and vibrational state distributions, and the probability amplitudes of these distributions are the targets for the NN model. Training data had collision energies between 5,000 K and 50,000 K at 5,000 K intervals, initial vibrational numbers in [0, 1, 3, 5, 7, 9] and initial rotational numbers in [0, 2, 10, 15, 20]. The probability amplitudes for final rotational numbers are predicted for every other number (e.g. $j_f$ = 0, 2, 4,$\ \ldots$). The probability amplitudes for intermediate final rotational quantum numbers (e.g. $j_f$ = 1, 3, 5,$\ \ldots$) can be obtained by a linear interpolation since the final state distributions are smooth and roughly locally linear ~\cite{Arnold2022}. Since less vibrational quanta were occupied, the targets for final vibrational distributions consisted of the probability amplitudes for quantum numbers $v_i$ = 0, 1, 2, ... m, with m depending on the particular isotopologue and process. Likewise, the maximum $j_f$ also varied depending on the isotopologue and process. Each NN model's output vector contains both the final rotational and vibrational distributions for its particular task, which is the combination of isotopologue and process it makes predictions for. 

\begin{table}[h]
    \centering
    \begin{tabular}{|c|c|c|c|c|}
        \hline
        Model & Layers & Neurons & Parameters & Training set size  \\
        \hline
        NNRH & 26 & 666 & 19277 & 242  \\
        \hline
        NNRD & 26 & 1027 & 53433 & 260 \\
        \hline
        NNRT & 26 & 1047 & 54153 & 260 \\
        \hline 
        NNIH & 26 & 666 & 19277 & 242 \\
        \hline
        NNID & 20 & 546 & 16832 & 260   \\
        \hline
        NNIso & 22 & 972 & 52573 & 502  \\
        \hline
    \end{tabular}
    \caption{The model requirements in terms of neural network architecture and training data set size. The required number of trajectories per model is on the order of $10^{7}$ since each training example requires 160000 trajectories.}
    \label{model_reqs}
\end{table}

 Given the need to predict multiple processes (reactive and inelastic) across three isotopologues of the reaction, a total of six models were developed, as shown in Table~\ref{model_reqs}. For each isotopologue, a final state distribution model was made for the reactive process. These three models from now on will be named NNRH, NNRD, and NNRT, standing for Neural Network for the Reactive process of the Hydrogen isotopologue, Neural Network for the Reactive process of the Deuterium isotopologue, and so on. For the hydrogen and deuterium isotopologues, models were developed to predict the inelastic process. These two models are named NNIH and NNID, standing for Neural Network for the Inelastic process of the Hydrogen isotopologue, and so on. Finally, one model was made for the purpose of predicting the reactive process for all three isotopologues simultaneously, and this model is named NNIso, or Neural Network for predicting Isotopic effects.

 The machine-learned neural network models used to predict the final ro-vibrational state distributions are feed-forward deep neural networks. The number of layers varies between models, and the various models' architectures, in terms of aggregates, are provided in Table \ref{model_reqs}. The number of neurons per layer is gradually increased layer by layer to get from the dimension of the input vector to the final dimension size of the target vector, which was always larger given the need to predict the probability amplitudes over a distribution of final quantum numbers. Indeed, this general pattern of expanding the NN width from the input to the output was found to be effective in previous work ~\cite{Arnold2022}. The activation function of all hidden layers except the last was the softplus function, given by
 \begin{equation}
     \text{softplus}(x) = \ln(1 + e^{x}). 
\end{equation}
 The final layer of each model, which produces the output vector of probability amplitudes, uses the sigmoid activation function since the targets are all probabilities and they, as well as their square roots, which were used for training, lie in [0, 1]. The Adam optimizer was used to fit the parameters of the NN models ~\cite{kingma2014adam}. The parameters were initialized using Glorot normal initialization ~\cite{pmlr-v9-glorot10a}. The learning rates of the models varied as the learning rate is a hyper-parameter of each individual model, but they were near 0.05. Tens of thousands of training epochs were used for each model with the total number overall ranging between approximately 45,000 - 75,000 epochs. The number of training epochs was of course another hyper-parameter allowed to vary between models. The mean squared deviation (MSD) loss function was used for the training process. In this work, the NN models were implemented using the TensorFlow package with Keras ~\cite{tensorflow2015-whitepaper,chollet2015keras}.

\begin{figure*}[t!]
    \includegraphics[width=1\linewidth]{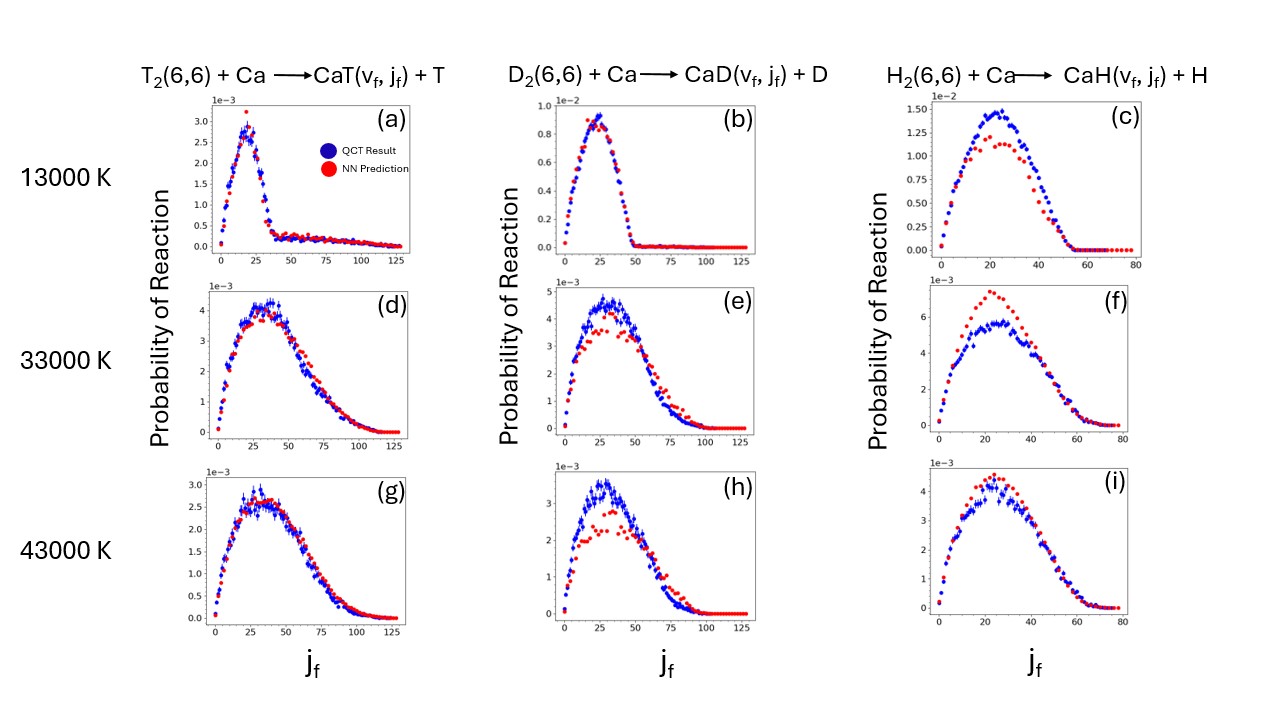}
    \caption{A sample of test results for the prediction of final reactive rotational state distributions for an initial state of (v$_i$, j$_i$) = (6, 6). The columns are organized by isotopologue; i.e., the first, second and third columns refer to the NNRT, NNRD and NNRH models, respectively. The rows are organized by the collision energy in K. 
    Points in blue indicate the results from QCT simulations and those in red indicate NN predictions.}
    \label{react_jf}
\end{figure*}

\subsection{Model evaluation}
The performance of the NN models was determined using test data generated for initial states that were not included in the training dataset. The performance accuracy of the models on the test data is calculated using the root mean square error (RMSE) for each test example, given as

\begin{equation}
\text{RMSE} = \sqrt{\frac{1}{N}\sum_{i=1}^{i=N} (y_i - \hat{y_i})^2},
\end{equation}
and it is calculated for each target final state distribution by summing over all $N$ NN predictions, $\hat{y_i}$, and comparing to QCT results, $y_i$. For each test example, which is a single final rotational or vibrational state distribution for a given initial state, the RMSE is calculated over all points for which NN predictions are available. In cases where the NN prediction exists but the QCT result is not specified, the QCT result is assumed to be 0 since the outcome in question was not observed in the QCT simulation and thus has an estimated probability of 0. The final reported error is then the ratio of the RMSE to the maximum probability amplitude for the distribution as given by the QCT calculation. The reported error is thus the relative RMSE

\begin{equation}
\label{relrmse}
\text{Relative \ RMSE} = \frac{\sqrt{\frac{1}{N}\sum_{i=1}^{i=N} (y_i - \hat{y_i})^2}}{\text{max}(\text{QCT \ distribution})},
\end{equation}
and represents how large the error is for a given predicted distribution relative to the maximum value available from the QCT result for that distribution. 

\section{Results}\label{results}

The NN model predictions for the product state distributions are divided into three categories. First, the NN models for reactive processes trained on a single isotopologue (NNRH, NNRD and NNRT). Second, for the NN models for inelastic collisions trained on a single isotopologue (NNIH and NNID), and third, for the NN model trained on two isotopologues of the reaction (NNiso), namely H$_2$ + Ca and T$_2$ + Ca, that then predicts the D$_2$ + Ca reaction in the testing data set.

\subsection{Product state distributions of reactive processes}

The three single reaction predictors (NNRH, NNRD, and NNRT) make predictions for the final rotational and vibrational state distributions of the reaction upon which they were trained. The test data for these models consisted of initial states ($v_i$, $j_i$, $E_c$) not included in the training data set. Here, we tested the models' ability to make predictions in the interpolation regime, meaning the initial states in the testing set are combinations of $v_i$, $j_i$, and $E_c$ that are within the range of the values of those variables presented in the training set. Given the small training data set size, see Table \ref{model_reqs}, the models are expected to have only limited extrapolation capability. 

\begin{figure*}[t!]
    \centering
    \includegraphics[width=\textwidth]{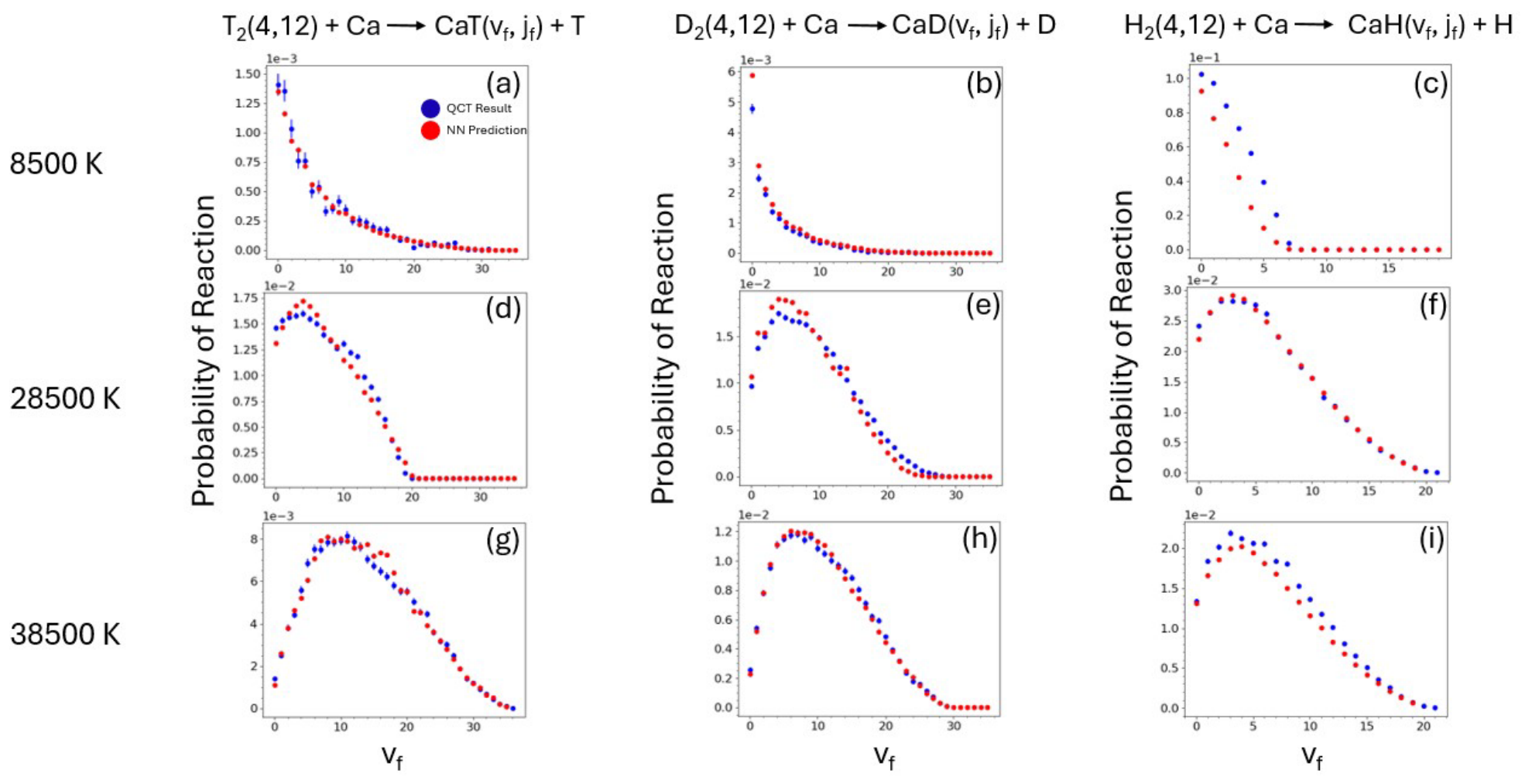}
    \caption{A sample of test results for the prediction of reactive final vibrational state distributions for an initial state of (v$_i$, j$_i$) = (4, 12). The columns are organized by isotopologue, i.e., first, second and third column refer to NNRT, NNRD and NNRH models, respectively. The rows are organized by the collision energy in K. 
    Points in blue indicate the results from QCT simulations and those in red indicate NN predictions.}
    \label{react_vf}
\end{figure*}

The results for the rotational product state distribution for Ca + H$_2(v_i,j_i)\rightarrow$ CaH$(v_f,j_f)$ + H, including all the isotopic variants of hydrogen -- deuterium and tritium -- are shown in Fig.~\ref{react_jf}. Specifically, the initial state is chosen as ($v_i$, $j_i$) = (6, 6), with collision energies of 13000~K, 33000~K and 43000~K. The figure shows a comparison between the NN predictions and the QCT calculations: panels (a), (d) and (g) shows the performance of the NNRT model; panels (b), (e) and (h) the performance of the NNRD model, whereas panels (c), (f) and (i) correspond to the NNRH model. The NN models agree extremely well with the QCT calculations, both qualitatively and quantitatively, despite the fact that, for deuterium and tritium, the largest probability is $\sim 10^{-3}$. As a result, the NN models are capable of describibing any subtlety of the rotational product state distribution.

In the case of the vibrational product state distribution, the results are shown in Fig.~\ref{react_vf}, where a remarkable agreement between the NN predictions and the QCT simulations is noticed. For instance, the NN models capture all the most relevant features: highly vibrational states are populated for higher collision energies. It is worth emphasizing that for these results, we used the same training set as for the ones in Fig.~\ref{react_jf}, but in this case the test set is different.

Based on the results displayed in Figs.~\ref{react_jf} and \ref{react_vf}, even with small training data sets, the models did not overfit to noise and instead learned the underlying relationships. In other words, our NN models capture the physics properly but ignore the possible small noise effects inherent to the Monte Carlo approach employed in the data generation through QCT simulations.

\begin{figure*}[t!]
    \centering
    \includegraphics[width=\linewidth]{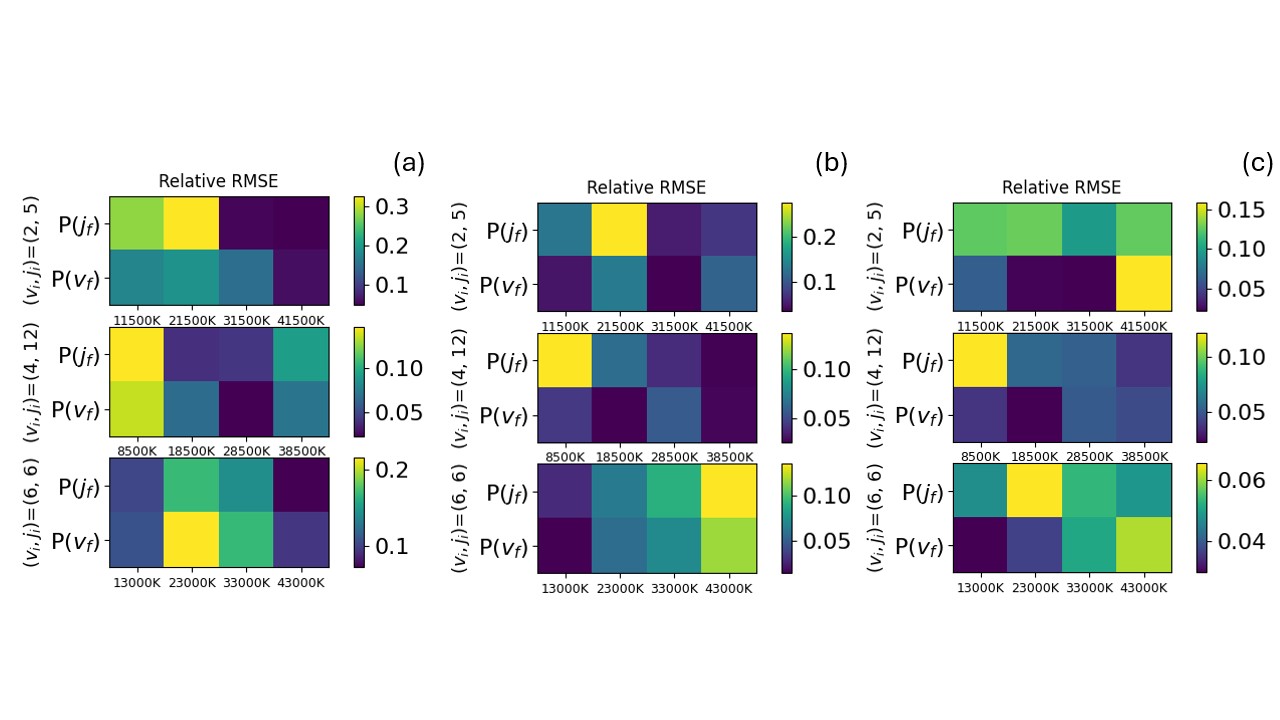}
    \caption{The relative RMSE values of the models trained on reactions. Panels a, b, and c are for NNRH, NNRD, and NNRT, respectively. In each panel, each heat map corresponds to a different initial ro-vibrational state (y-axis label), rows correspond to the distribution, either rotational $P(j_f)$, or vibrational $P(v_f)$, and columns correspond to the collision energies in K.}
    \label{react_err_maps}
\end{figure*}


Fig. \ref{react_err_maps} presents a general assessment of the performance of the single reaction predictors. This figure displays a heat map of the relative RMSE for the product state distributions (see Eq.~\ref{relrmse}), including vibrational and rotational product state distributions. First, one notices a rather complex pattern in the error distribution. For instance, the model NNRH shows a relative RMSE below 0.2 for initial states $(v_i, \ j_i)=(6, \ 6)$ and $(4, \ 12)$, independently of the collision energy. On the contrary, the same model for $(v_i, \ j_i)=(2, \ 5)$, shows at high collision energies a relative RMSE of 0.3. The same applies to the NNRD and NNRT models. It is worth noting the remarkable performance of the NNRT model for $(v_i, \ j_i)=(6, \ 6)$, showing a relative RMSE below 0.06 independent of the collision energy. However, independent of the complex pattern of the error distribution, the overall combined error of the model for the rotational and vibrational state distributions is less than 0.15, or 15\%.

\subsection{Product state distributions of inelastic processes}

 We have developed two NN models to predict the final rotational and vibrational state distributions for the inelastic collisions between hydrogen or deuterium molecules with calcium, labeled as NNIH and NNID, respectively, as shown in Table~\ref{model_reqs}. These two models each predict the final rotational and vibrational distributions for their respective isotopologues. Additionally, both models use the same featurization as that used for the single reaction predictors. 
 
 \begin{figure}[h]
    \centering
    \includegraphics[width=1.3\linewidth]{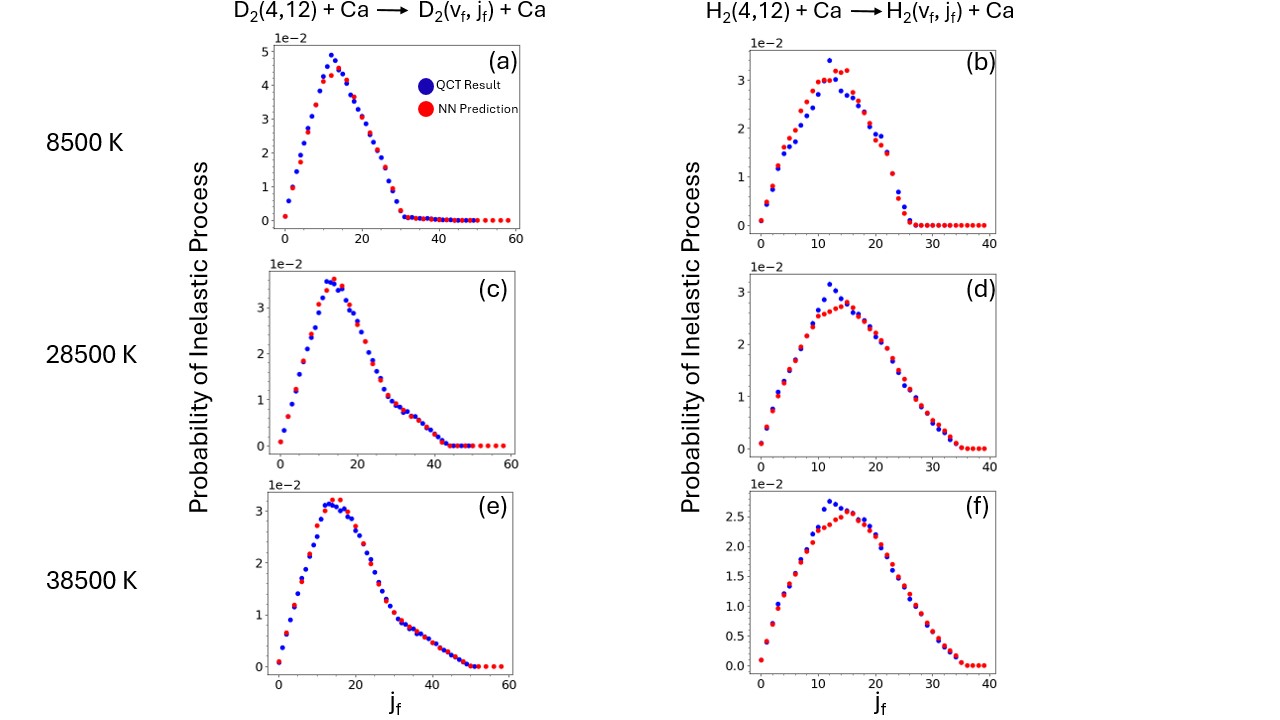}
    \caption{A sample of test results for the prediction of final rotational state distributions of the inelastic process for an initial state of (v$_i$, j$_i$) = (4, 12). The columns are organized by isotopologue, and the rows are organized by the collision energy in K. Points in blue indicate the results from QCT simulations and those in red indicate NN predictions. Panels (a), (c) and (e) refer to NNRD, and panels (b), (d) and (f) are for the NNRH model. }
    \label{quench_jf}
\end{figure}

\begin{figure}[h!]
    \centering
        \includegraphics[width=1.4\linewidth]{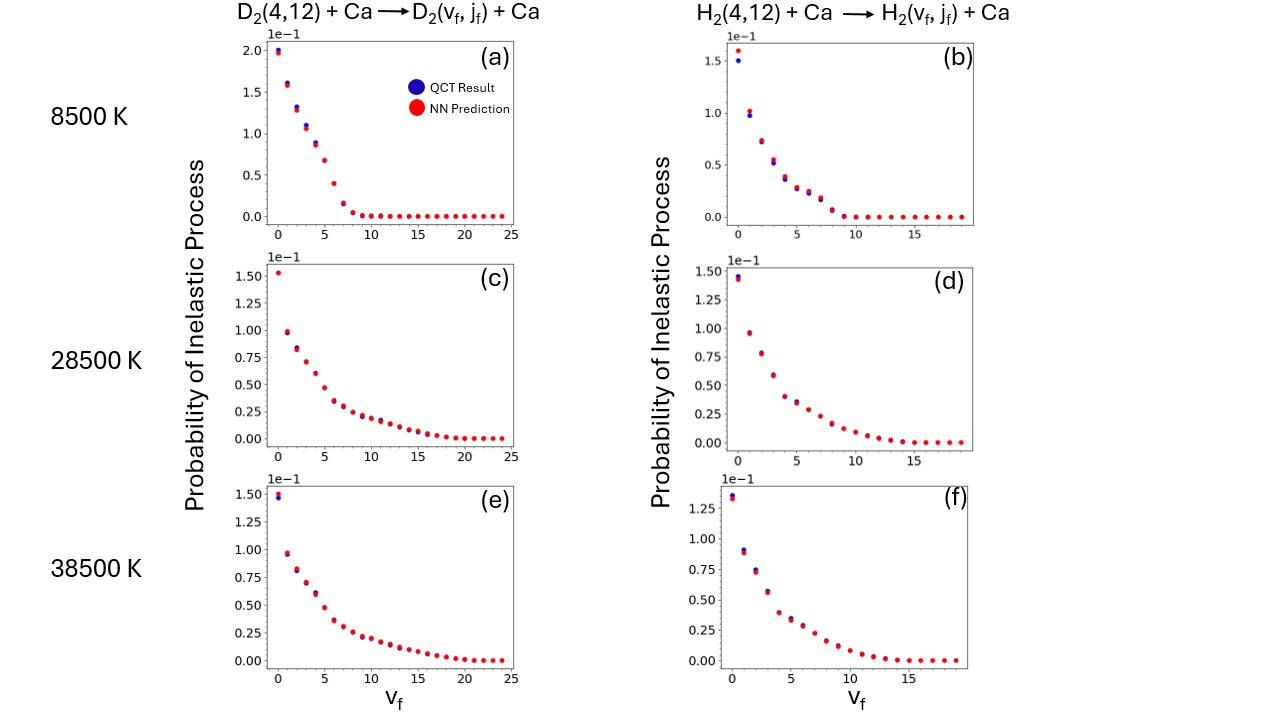}
    \caption{A sample of test results for the prediction of final vibrational state distributions of the inelastic process for an initial state of (v$_i$, j$_i$) = (4, 12). The columns are organized by isotopologue, and the rows are organized by the collision energy in K. Points in blue indicate the results from QCT simulations and those in red indicate NN predictions.}
    \label{quench_vf}
\end{figure}

 A sample of test results for the final state rotational distribution of the inelastic collision models is provided in Fig. \ref{quench_jf}. In this case, as in the case of reactions, the NN models reproduce remarkably well the QCT calculations. As a result, we can conclude that the NN models are learning the underlying chemistry properly. Similarly, in Fig.~\ref{quench_vf}, we present our results for the final vibrational product state distribution. The performance of the NNs is outstanding independent of the collision energy and final vibrational state. Therefore, as with the reactive process predictors, the inelastic predictors demonstrate a capability to make predictions on test data in the interpolation regime that succeed both in terms of numerical accuracy and the shapes of the distributions.

\begin{figure}[h!]
    \centering
    \includegraphics[width=\linewidth]{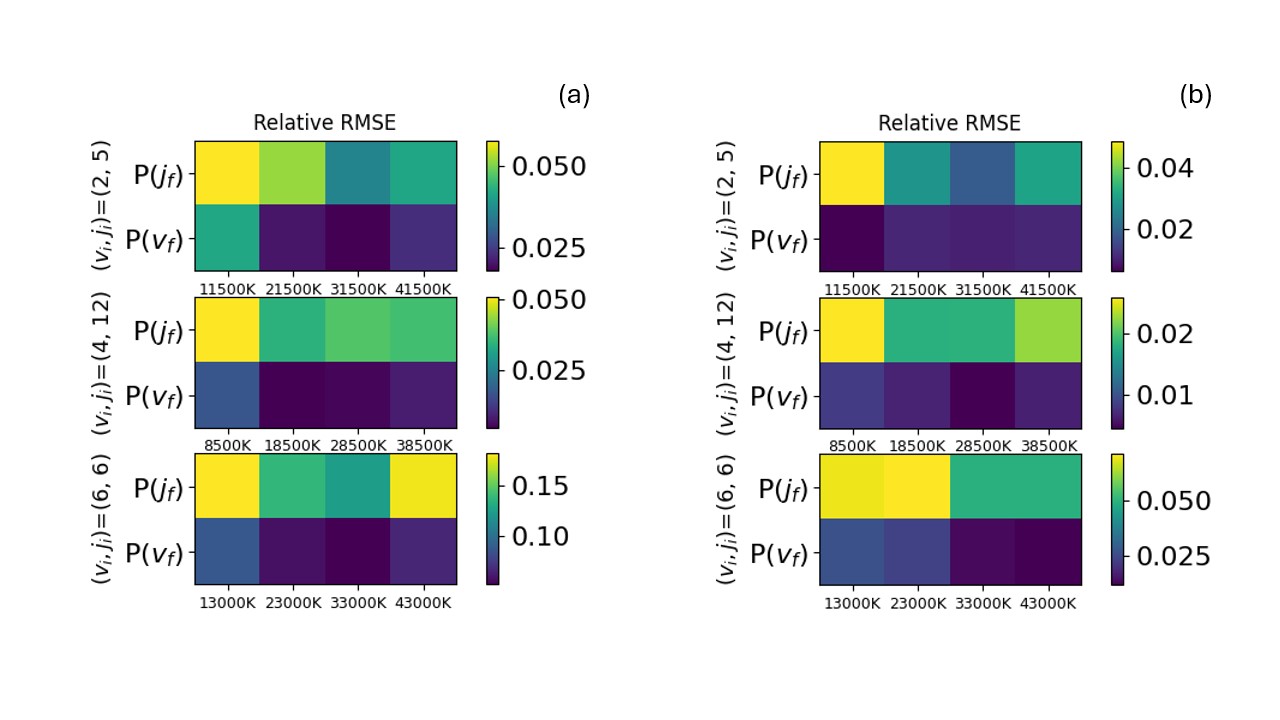}
    \caption{The relative RMSE values of the models trained on inelastic processes. Panels a and b are for NNIH and NNID, respectively. In each panel, each heat map corresponds to a different initial ro-vibrational state (y-axis label), rows correspond to the distribution, either rotational $P(j_f)$, or vibrational $P(v_f)$, and columns correspond to the collision energies in K.}
    \label{quench_err_maps}
\end{figure}

  The relative RMSE values for the test results of the inelastic collision models are provided in Fig.~\ref{quench_err_maps}. First, we notice that the relative errors are smaller than the models for reactions. In this case the relative errors do not exceed 15\%. There are even some cases where the relative error is 2.5\%, showing excellent performance. Therefore, it is clear that the inelastic collision models are capable of predicting the shapes of the final state distributions with high fidelity alongside high numerical accuracy. Using the same featurization as the single reaction predictors, it is clear that neural network models can learn the relationship between the initial state space and the final rotational and vibrational state distributions with high accuracy. It is apparent that the inelastic predictors outperform their reactive process counterparts for most if not all of the test examples provided in the testing data set. Thus, the featurization used for the reactive and inelastic models is universal between processes.

Clearly, the inelastic models outperform their reactive counterparts for nearly all test examples. This can be explained by the fact that reactions are less frequent than inelastic processes for the range of collision energies explored in this work, yielding smaller probabilities in comparison with inelastic processes. Hence, the NN models for reactions have a harder time capturing the underlying reaction mechanism. 

\begin{figure*}[t!]
    \centering
    \includegraphics[width=\linewidth]{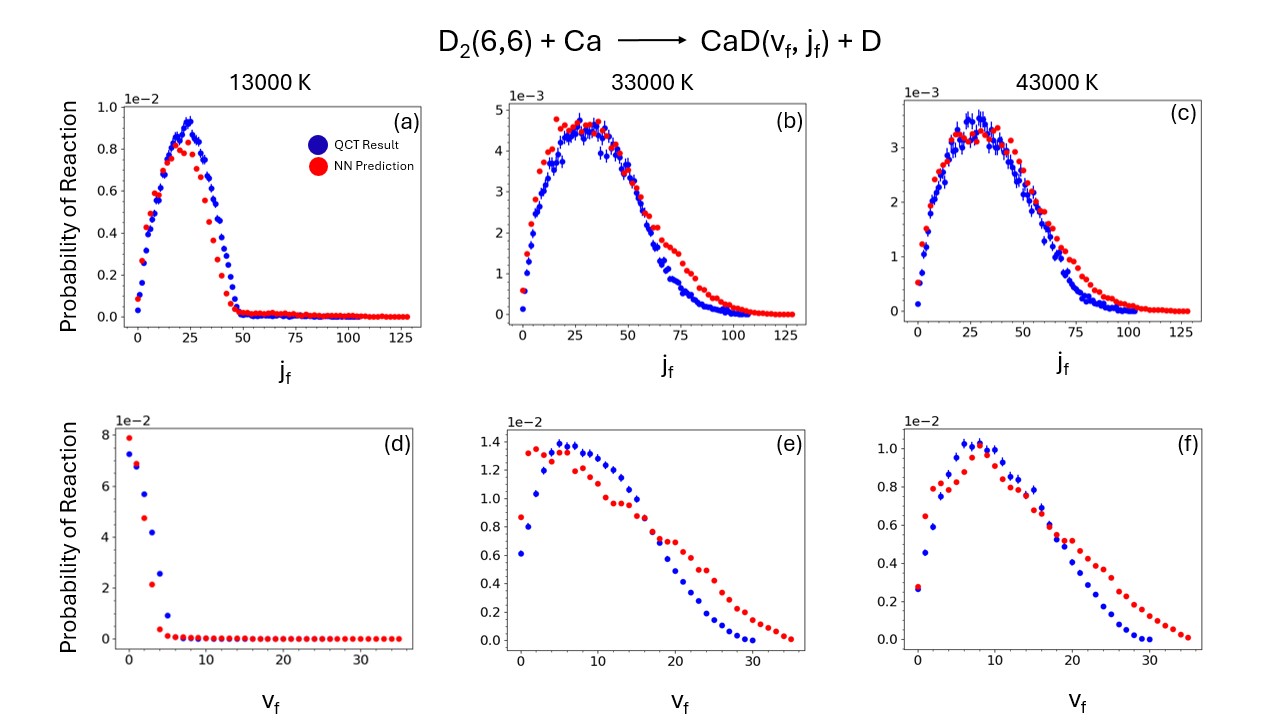}
    \caption{A sample of test results for the prediction of final reactive state distributions for an initial ro-vibrational state of (v$_i$, j$_i$) = (6, 6) of the D$_2$ + Ca $\rightarrow$ CaD +D reaction. The columns are organized by collision energy, and all results are for predicting the deuterium reaction. The model that produced these predictions was trained on just the hydrogen and tritium reactions. Points in blue indicate the results from QCT simulations and those in red indicate neural network predictions.}
    \label{isotopic_effects}
\end{figure*}

The training data set size between each inelastic model and their reactive counterpart (same isotopologue) are equivalent, and these training sets show the same portions of the initial state space. The inelastic models do not capture the final state distributions perfectly, so there is the possibility that increasing the training data set size could improve model performance. However, even with a small data set it is clear that the inelastic process for the hydrogen and deuterium collisions with calcium can be effectively predicted by a simple feed-forward neural network.

\subsection{Prediction results for a model trained on multiple reactions}

Until now, the training data contained the same chemical species as the final product state distribution; i.e., the model is exposed to the same reaction we want to predict. This is the workhorse of many ideas on machine learning prediction models for atom-molecule collisions~\cite{Arnold2020,Arnold2022,COO,Dissociation,Koner2019,Panesi2023,Veliz2022}. However, to develop a universal predictor, it would be necessary to make predictions of a reaction across the chemical space. In other words, the training data could be different from the reaction that has to be predicted, which is a very demanding property. However, Ca + H$_2$ collisions serve as a candidate to explore the universality of machine learning models for chemical reactions since it could be easily generalized, including hydrogen isotopologues, and hence, the same reaction can lead to three different ones. In this case, the hydrogen isotopes only enlarge the chemical space. However, as shown below, the isotopic effects on the reaction are substantial. On the other hand, it is worth mentioning that the predictive power of these models could be compromised by the zero-point energy effects characteristic of a full quantitative treatment and absent in a QCT approach.

Our model NNIso is trained on data from Ca + H$_2(v_i,j_i)\rightarrow$ CaH$(v_f,j_f)$ + H and Ca + T$_2(v_i,j_i)\rightarrow$ CaT$(v_f,j_f)$ + T, but it predicts the product state distribution for D$_2(v_i,j_i)\rightarrow$ CaD$(v_f,j_f)$ + D. The NN model has never been exposed to any deuterium containing reaction, and still it will predict the outcome of the reaction. To do so, as previously mentioned, this model was given an expanded feature vector that captures variables related to the isotopic differences in the chemical system. With this expanded featurization, this model generalizes between the isotopologues of the reaction. The capability of this model to produce accurate results indicates that it has effectively learned the isotopic effects of the reaction. It should be noted that isotopic variants of the reaction other than those mentioned exist, such as where the reactant diatom is composed of two atoms of different hydrogen isotopes. For these particular isotopic variants, our methodology could be extended to predict these reactions by developing more models or enlarging current models to accommodate two distinguishable reactive channels, one for each possible product diatom. However, this is beyond the scope of this work as our models currently are used to predict a single reactive channel of an atom-diatom collision. 
 
\begin{figure}[h]
    \includegraphics[width=1\linewidth]{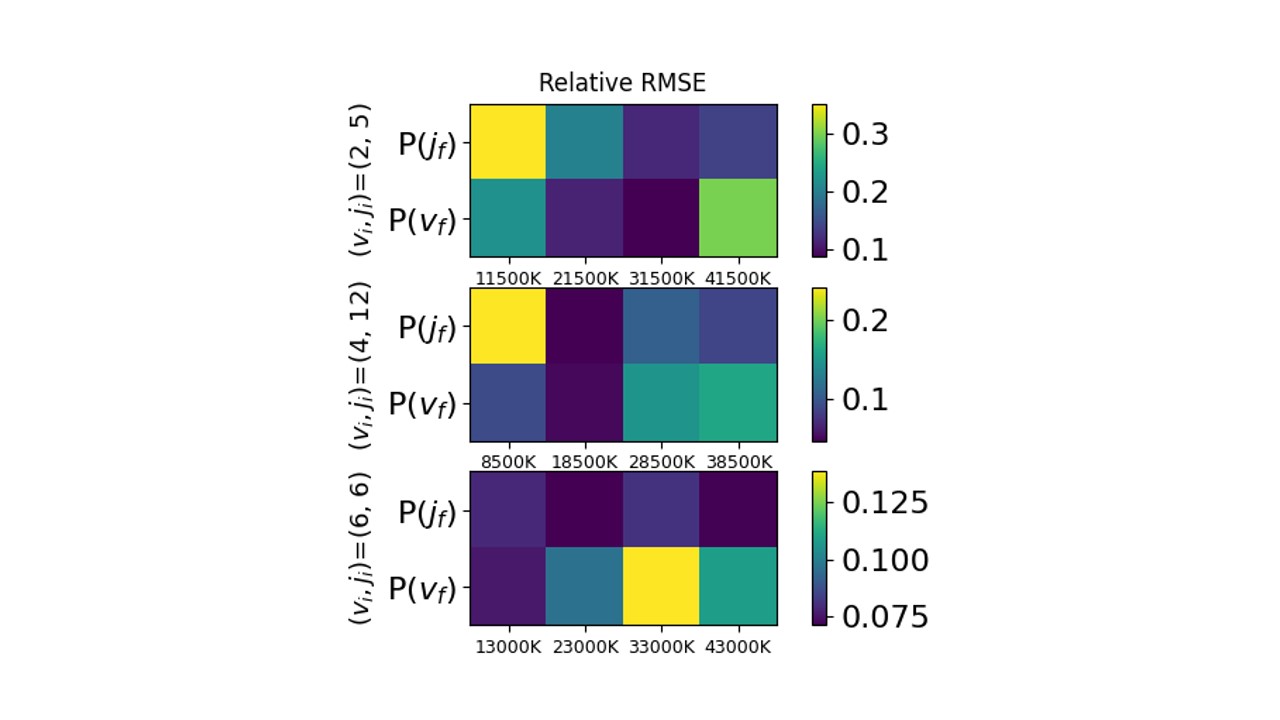}
    \caption{The relative RMSE values for the test data for predicting isotopic effects in reactions (model NNIso). The model evaluated was trained on the hydrogen and tritium reactive processes and the test data consists of deuterium reactive collisions. Each heat map corresponds to a different initial ro-vibrational state (y-axis label), rows correspond to the distribution, either rotational $P(j_f)$, or vibrational $P(v_f)$, and columns correspond to the collision energies in K.}
    \label{iso_err_map}
\end{figure}

The test performance of this model is captured in Figs.~\ref{isotopic_effects} and \ref{iso_err_map}, with the latter providing the relative RMSE values. From Fig.~\ref{isotopic_effects}, which shows the  product state distributions, it is clear that the model is capable of capturing the shape of the final rotational and vibrational state distributions. The model predictions still follow the trend present in the QCT results and capture the shapes of the actual distributions. Additionally, this model is capable of making numerically accurate predictions, as evidenced by the overlap of the predicted and actual distributions in panels (a)-(c) of Fig.~\ref{isotopic_effects}. Thus, this model is capable of generalizing between isotopologues of the reaction to make effective predictions of the final state distributions for members of the test data set. 

Though this model that generalizes between isotopologues has shown success in predicting members of the testing sample, it struggles more than the single reaction predictor counterparts to make equally accurate predictions. This result arises despite NNIso being given more training data than any individual single reaction predictor (NNIso was trained on both the hyrdrogen and tritium isotopologues). However, it should be noted that the multiple reaction predictor saw the same regions of the initial state space, albeit between two different reactions. So, while more training data was provided to the multiple reaction predictor, more was needed to compensate for the model needing to learn the isotopic effects with the extended featurization. If the model were provided with some deuterium training examples, it could better generalize between the isotopologues, but we sought to push the model by having it predict a reaction not previously included in the training data set. The featurization for this model could be improved to allow the neural network to learn the isotopic effects better. While the original featurization used for the previous models has been shown to produce effective predictions, it is still redundant (e.g., by including both angular momentum and rotational energy) and perhaps removing this redundancy would allow for better training of the model to learn the isotopic effects. Also, there could be yet a better-extended featurization that could be used to inform the model about how to predict the isotopic effects of the reactions.

\section{Conclusions}\label{conclusions}

Our work explores the capabilities of deep learning in predicting atom-molecule reaction outcomes. Specifically, we focus on Ca + H$_2$ collisions in the hyper-thermal regime as a prototypical example, including isotopologue variations. Using a deep feed-forward neural network, we demonstrate it is possible to use a single featurization to learn the final reactive ro-vibrational state distributions across isotopologues of an atom-diatom reactive collision. Additionally, we show that this featurization is general enough to allow a deep neural network to learn either the reactive or inelastic process. 

To expand on this, we develop an NN model capable of learning isotope variants of a given reaction; i.e., the model is capable of predicting the outcome of a reaction using information from other reactions. By augmenting the featurization to include information about the masses of the system and select spectroscopic constants of the reactant diatom, we demonstrate it is possible for a single NN model to learn all of the hydrogen isotopic variants of the reaction H$_2$ + Ca $\rightarrow$ CaH + H. Efforts such as these to design machine-learned neural networks to generalize between more reactions will move us closer to a universal model for atom-diatom reactive collisions.

\section*{acknowledgements}
This work was supported by the United States Air Force Office of Scientific Research [grant number FA9550-23-1-0202]. The authors thank Prof. M. Meuwly for reading the manuscript and for fruitful discussions.

\bibliography{apssamp}

\end{document}